\documentclass[aps,pre,reprint,showpacs,showkeys,amsmath]{revtex4-1}
\usepackage{hyperref}
\usepackage{graphicx}

\begin{document}

\title{Domain-wall complexes in 1D ferromagnets and critical media}
\author{Andrzej Janutka}
\email{Andrzej.Janutka@pwr.wroc.pl}
\affiliation{Institute of Physics, Wroclaw University of Technology, Wybrze\.ze Wyspia\'nskiego 27, 50-370 Wroc{\l}aw, Poland}

\begin{abstract}
Interactions of domain walls (DWs) are analyzed with relevance to formation of stationary bubbles (complexes of two DWs) 
 and complexes of many domains in one dimensional systems. I investigate the domain structures in ferromagnets which are described with 
 the Landau-Lifshitz equation as well as the domains in critical systems described with the Ginzburg-Landau equation.
 Supplementing previous author studies on the creation of hard bubbles [formed by one Bloch DW and one Neel (Ising) DW] 
 in the presence of an external (magnetic) field, the soft bubbles consisting of two Bloch DWs or two Neel (Ising) DWs are
 studied in detail. The interactions of two DWs of the same kind are studied in the framework of a perturbation calculus. 
\end{abstract}


\maketitle
\newpage

\section{Introduction}

Localized and patterned (labirynth) structures in bistable media are widely considered with relevance to the storage of
 binary information \cite{cou04,mik06}. Far enough from the phase-transition point, the complexes of 
 DWs are observed in chemical reactors, magnetic and polar (solid and liquid) media while their basic properties 
 are described with equations of motion with are similar for different media. In particular, far from the
 critical regime, the magnetization (polarization) dynamics is described with the Landau-Lifshitz-Gilbert (LLG) equation
 while near the criticality it is governed by the Ginbzurg-Landau (GL) equation. Recently, one observes an especial interest 
 in 1D complexes of magnetic and polar DWs due to hopes for miniaturization of memory carriers which result 
 from achievements of current nanowire technology \cite{par08,hay08,sco07}. The problem of stability
 of many-domain complexes is connected to the need of switching the domain-encoded binary information 
 on demand since such process induces unbalanced interactions of the DWs.  

In the present work, I analyze binary interactions between the DWs (of the Bloch and Neel-Ising type) 
 with dependence on the distance of their separation and their chiralities (opposite or like) and 
 I study the formation of magnetization bubbles in 1D far from and near the criticality. 
 It enables me to consider the stability of many-domain structures. 
 With correspondence to 2D magnetic bubbles (which are called hard ones when their boundary is composed of alternating 
 Bloch and Neel lines), I divide the 1D bubbles into hard ones (composed of one Bloch and one Neel-Ising DWs)
 and soft ones (composed of two Bloch DWs or two Neel-Ising DWs) \cite{ros72,lee80}. In my recent papers on the externally-driven
 motion of DWs, I have studied the formation of 1D hard bubbles via collision of a Bloch DW with a Neel (Ising) DW in the normal 
 and critical  regimes, (solving the LLG and GL equations, respectively) \cite{jan11,jan11a}. Here, the energy of soft bubbles
 is studied as function of the bubble length within a perturbation calcus in order to complete the
 picture of the DW interactions. I follow a perturbation approach to the DW interaction developed in Ref. \cite{bar07}
 with relevance to the parametrically-driven nonlinear Schrodinger equation. It differs from previous ones (e.g. \cite{kor92})
 in terms of the perturbation expansion of the dynamical parameter (magnetization) whose present form ensures conservation 
 of the magnetization length. 
 
With application to ferromagnets, the 1D approximation is relevant to crystalline nanowires with strong bulk anisotropy
 compared to surface magnetostatic effects (and, especially, with circular cross section). The present analysis provides
 a basis for comparison of the DW interactions in such magnets to the interactions of DWs in the nanostripes 
 of noncrystalline ferromagnets which are studied in Ref. \cite{jan12}. 
 
Binary interactions od the DWs and the stability of 1D bubbles are considered in sections II and III, with relevance to the ferromagnetic
 wire at low temperatures and to the subcritical systems, respectively. In section IV, the stability of many-DW 1D systems is considered
 on the basis of previous-section results.

\section{Magnetic bubbles at low temperatures}

I consider DW solutions to the LLG equation  
\begin{eqnarray}
\frac{\partial{\bf m}}{\partial t}=\frac{J}{M}{\bf m}\times\frac{\partial^{2}{\bf m}}{\partial x^{2}}
+\gamma{\bf m}\times{\bf H}+\frac{\beta_{1}}{M}({\bf m}\cdot\hat{i}){\bf m}\times\hat{i}
\nonumber\\
-\frac{\beta_{2}}{M}({\bf m}\cdot\hat{j}){\bf m}\times\hat{j}
-\frac{\alpha}{M}{\bf m}\times\frac{\partial{\bf m}}{\partial t}.
\label{LLG}
\end{eqnarray}
The first term of the r.h.s. of (\ref{LLG}) relates to the exchange spin interaction while the second (Zeeman) term
 depends on the external magnetic field ${\bf H}=(H_{x},0,0)$, thus, $\gamma$ denotes the giromagnetic factor. 
 The constant $\beta_{1(2)}$ determines strength of the easy axis (plane) anisotropy and
 $\hat{i}\equiv(1,0,0)$, $\hat{j}\equiv(0,1,0)$. I notice that the long axis of a nanowire is an easy axis for
 the most often investigated magnets, however, another choice of the anisotropy axes 
 does not influence the magnetization dynamics. Since LLG equation is valid only when the constraint $|{\bf m}|=M$ 
 is satisfied, one writes equivalent to (\ref{LLG}) equations of the unconstrained dynamics. 
 Introducing $m_{\pm}\equiv m_{y}\pm{\rm i}m_{z}$, one represents 
 the magnetization components using a pair of complex functions $g(x,t)$, $f(x,t)$. The relation between 
 the primary and secondary dynamical variables  
\begin{eqnarray}
m_{+}=\frac{2M}{f^{*}/g+g^{*}/f},
\hspace*{2em}
m_{x}=M\frac{f^{*}/g-g^{*}/f}{f^{*}/g+g^{*}/f}
\label{transform}
\end{eqnarray}
ensures that $|{\bf m}|=M$. Inserting (\ref{transform}) into (\ref{LLG}) leads,
 following the Hirota method for solving nonlinear equations \cite{bog80,kos90}, to 
\begin{eqnarray}
f\left[-{\rm i}D_{t}+JD_{x}^{2}+\alpha D_{t}\right]f^{*}\cdot g
+Jg^{*}D_{x}^{2}g\cdot g&&
\nonumber\\
-\left(\gamma H_{x}+\beta_{1}+\frac{\beta_{2}}{2}\right)|f|^{2}g
-\frac{\beta_{2}}{2}f^{*2}g^{*}&=&0,
\nonumber\\
g^{*}\left[-{\rm i}D_{t}-JD_{x}^{2}+\alpha D_{t}\right]f^{*}\cdot g
-JfD_{x}^{2}f^{*}\cdot f^{*}&&
\nonumber\\
+\left(-\gamma H_{x}+\beta_{1}+\frac{\beta_{2}}{2}\right)|g|^{2}f^{*}
+\frac{\beta_{2}}{2}g^{2}f&=&0,
\label{secondary-eq}
\end{eqnarray}
where $D_{t}$, $D_{x}$ denote Hirota operators of differentiation which are defined by 
\begin{eqnarray}
D_{t}^{m}D_{x}^{n}b(x,t)\cdot c(x,t)\equiv
(\partial/\partial t-\partial/\partial t^{'})^{m}
\nonumber\\
\times
(\partial/\partial x-\partial/\partial x^{'})^{n}b(x,t)c(x^{'},t^{'})|_{x=x^{'},t=t^{'}}.
\nonumber
\end{eqnarray}
For ${\bf H}=0$, the stationary single-DW solutions to (\ref{secondary-eq}) take the form
\begin{eqnarray}
f=1,\hspace*{2em}g=w{\rm e}^{kx},\hspace*{2em}k=k^{*},
\label{single_DW}
\end{eqnarray}
where $|k|=\sqrt{\beta_{1}/J}$, $w=-w^{*}$, (a Bloch DW)
 or $|k|=\sqrt{(\beta_{1}+\beta_{2})/J}$, $w=w^{*}$, (a Neel DW). Let us define $\varphi$
 following ${\rm e}^{{\rm i}\varphi}=w/|w|$, thus, $\varphi=\pm\pi/2$ for Bloch DW while
 $\varphi=0,\pi$ for Neel DW.
  
 
Since neither exact double-Bloch nor double-Neel solutions to the LLG equation are not known for the case 
 of zero external field, (the length of the soft bubbles diverges with $H_{x}\to 0$, \cite{bra94,eve10,sha06}), I analyze the interactions 
 of the relevant DW pairs (the pair of Bloch DWs and the pair of Neel DWs) within a perturbation calculus. Locally, 
 in the vicinity of the center of $j$th DW, ($j=1,2$), one can write the magnetization in the form
\begin{eqnarray} 
{\bf m}(x,0)={\bf m}^{(j)}(x)+\delta{\bf m}^{(j)}(x),
\end{eqnarray}
where ${\bf m}^{(j)}$ denotes a stationary single-DW solution to ($\ref{LLG}$) [which corresponds to (\ref{single_DW})]
\begin{eqnarray} 
m_{+}^{(j)}(x)&=&M{\rm e}^{{\rm i}\varphi_{j}}{\rm sech}[\sigma_{j}k(x-x_{0j})],
\nonumber\\
m_{x}^{(j)}(x)&=&-M{\rm tanh}[\sigma_{j}k(x-x_{0j})]
\label{profile1}
\end{eqnarray}
while $\delta{\bf m}^{(j)}$ denotes a perturbation due the presence of another DW. 
 Here $\sigma_{1}=-\sigma_{2}=\pm 1$, $\varphi_{1}=\pm\varphi_{2}$ and $|\varphi_{j}|=\pi/2$ (a pair of Bloch DWs)
 or $\varphi_{j}=0,\pi$ (a pair of Neel DWs). When assume $k>0$, $\sigma_{j}=1$ relates to the head-to-head spin
 structure while $\sigma_{j}=-1$ to the tail-to-tail structure.  
 In order to satisfy the constraint $|{\bf m}|=M$, I apply the perturbation of the form
\begin{eqnarray}
\delta{\bf m}^{(j)}&\approx&\left(\pm\frac{m_{x}^{(k)}}{M}-1\right){\bf m}^{(j)}
\pm\frac{m_{x}^{(j)}}{M}\left(0,m_{y}^{(k)},m_{z}^{(k)}\right)
\nonumber\\
&&\mp\frac{1}{M}\left(m_{y}^{(k)}m_{y}^{(j)}+m_{z}^{(j)}m_{z}^{(k)},0,0\right),
\end{eqnarray}
where $k\neq j$. It leads to 
\begin{eqnarray}
m_{x}&\approx&\pm\frac{1}{M}\left[m_{x}^{(1)}m_{x}^{(2)}-\frac{1}{2}\left(m_{+}^{(1)}m_{-}^{(2)}+m_{+}^{(2)}m_{-}^{(1)}\right)\right],
\nonumber\\
m_{+}&\approx&\pm\frac{1}{M}\left(m_{+}^{(1)}m_{x}^{(2)}+m_{+}^{(2)}m_{x}^{(1)}\right).
\label{lowest_order_LLG}
\end{eqnarray}
Plus and minus relate to the bubble magnetization parallel and antiparallel to $x$-axis, respectively. 
 This form of the initial magnetization reflects the fact that the interaction of topological solitons (DWs) is not simply
 due to their overlap while it is accompanied by some reorientation of whole the separating them domain.
 Therefore, the magnetization (\ref{lowest_order_LLG}) contains only products of the components of ${\bf m}^{(1)}$
 and ${\bf m}^{(2)}$ while it does not contain linear in ${\bf m}^{(j)}$ terms. Let $a\equiv k(x_{02}-x_{01})$. 
 Inserting (\ref{lowest_order_LLG}) into the Hamiltonian 
\begin{eqnarray}
{\cal H}={\cal H}_{0}+{\cal H}_{Z}&=&\frac{J}{2M}\bigg|\frac{\partial{\bf m}}{\partial x}\bigg|^{2}
+\frac{\beta_{1}}{2M}\left[M^{2}-({\bf m}\cdot\hat{i})^{2}\right]
\nonumber\\
&&+\frac{\beta_{2}}{2M}({\bf m}\cdot\hat{j})^{2}-\gamma{\bf H}\cdot{\bf m},
\end{eqnarray}
where ${\cal H}_{Z}$ denotes the Zeeman part of the Hamiltonian,
 one arrives at the dependence of the energy $E_{0(Z)}\equiv\int_{-\infty}^{\infty}{\cal H}_{0(Z)}{\rm d}x$
 on the distance between the DW centers
\begin{eqnarray}
E_{0}(a)&=&\frac{M(\beta_{1}+\theta\beta_{2})}{2k}\left[I_{1}^{\pm}(a)+I_{2}^{\pm}(a)\right]
\nonumber\\
&=&\frac{M\sqrt{J(\beta_{1}+\theta\beta_{2})}}{2}
{\rm csch}^{2\mp 1}(a/2)
\nonumber\\
&&\times{\rm sech}^{2\pm 1}(a/2)[-2a+{\rm sinh}(2a)],
\\
E_{Z}(a)&=&-\gamma H_{x}\frac{M}{k}I_{3}^{\pm}(a)
\nonumber\\
&=&-2\frac{\gamma H_{x}M\sqrt{J}}{\sqrt{\beta_{1}+\theta\beta_{2}}}
a[{\rm coth}(a)\mp{\rm csch}(a)],
\end{eqnarray}
where $\theta=1$ for Neel DWs while $\theta=0$ for Bloch DWs, and $I_{k}^{\pm}(a)$ denote integrals given in Appendix. 
 The upper signs correspond to the pair of the DWs of the opposite chiralities, the case $\varphi_{1}=\varphi_{2}$, while
 the lower signs correspond to the pair of the DWs of like chiralities, the case $\varphi_{1}=\varphi_{2}\pm\pi$.
 According to the plot of energy of the DW pair (Fig. 1), in absence of the external field, the interaction 
 is attractive when both the DWs are of opposite chiralities while it is repulsive in the case of like chiralities. 

\begin{figure}
\unitlength 1mm
\begin{center}
\begin{picture}(175,40)
\put(0,-5){\resizebox{85mm}{!}{\includegraphics{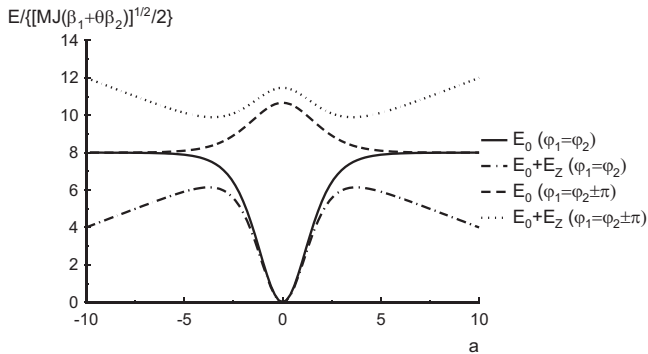}}}
\end{picture}
\end{center}
\caption{Energy of a pair of DWs as a function of the distance of their separation; solid (dash-dotted)
line - DWs of opposite chiralities (an untwisted pair) in absence (presence) of an external field, 
dashed (dotted) line - DWs of the same chiralities (a twisted pair) in absence (presence) of an external field.}
\end{figure}

In both the cases (of opposite and like chiralities), it is possible to create a stationary magnetization bubble
 applying an external magnetic field in the direction pursuant or opposite to the bubble magnetization, respectively.
 However, only in the case of $\varphi_{1}=\varphi_{2}\pm\pi$, the function $E(a)\equiv E_{0}(a)+E_{Z}(a)$ has a minimum
 at $a\neq 0$, (see Fig. 1). Therefore, only the bubble created by the DWs of like chiralities is stable against external
 perturbations. This stable bubble corresponds to the final state of the long-term evolution of a breather. 
 
For ${\bf H}\neq 0$, one can find strict double-Bloch and double-Neel solutions to LLG equation, 
 thus, one can verify our prediction on existence of soft stationary bubbles. 
 Assuming the stationary double-Bloch DW solution and double-Neel DW solution to (\ref{secondary-eq})
 to be of the form $f=1+v$, $g=w{\rm e}^{kx}+w{\rm e}^{-kx}$, $v=v^{*}$, one finds 
\begin{eqnarray}
|k|&=&\sqrt{\frac{\beta_{1}+\theta\beta_{2}-\gamma H_{x}}{J}},
\nonumber\\
v&=&-1\pm2\sqrt{\frac{\beta_{1}+\theta\beta_{2}}{\gamma H_{x}}-1}|w|,
\label{untwisted}
\end{eqnarray}
for $w=w^{*}$, $\theta=1$ (an untwisted double-Neel DW),
 and for $w=-w^{*}$, $\theta=0$ (an untwisted double-Bloch DW).
 Inserting $f=1+v$, $g=w{\rm e}^{kx}-w{\rm e}^{-kx}$ into (\ref{secondary-eq}), one finds
\begin{eqnarray}
|k|&=&\sqrt{\frac{\beta_{1}+\theta\beta_{2}-\gamma H_{x}}{J}},
\nonumber\\
v&=&-1\pm2\sqrt{1-\frac{\beta_{1}+\theta\beta_{2}}{\gamma H_{x}}}|w|,
\label{twisted}
\end{eqnarray}
for $w=w^{*}$, $\theta=1$ (a twisted double-Neel DW), and for
 $w=-w^{*}$, $\theta=0$ (a twisted double-Bloch DW). The untwisted double-DWs are called nuclei while the twisted double-DWs 
 are called $2\pi$-DWs. Since typically $\beta_{1}\gg\beta_{2}$ while $|H_{x}|=\beta_{1}/\gamma$ corresponds to the coercion-field 
 value, (hence, I assume $|H_{x}|\ll\beta_{1}/\gamma$), the untwisted and twisted double-DWs relate to different (parallel
 or antiparallel to the bubble magnetization, respectively) directions of the external magnetic field, which ensures 
 that $v$ in (\ref{untwisted}) or (\ref{twisted}) is determined. In order that the existence of nuclei was possible, 
 the DWs of like chiralities must attract, thus, the nucleus vanishes after turning the magnetic filed off.
 The prediction on stability of the $2\pi$-DWs and instability of nuclei is in accordance with a study of linear-excitation
 scattering on the soft magnetization bubbles and with simulations \cite{bra94,sha06}. 
  
\begin{figure}
\unitlength 1mm
\begin{center}
\begin{picture}(175,30)
\put(0,-5){\resizebox{85mm}{!}{\includegraphics{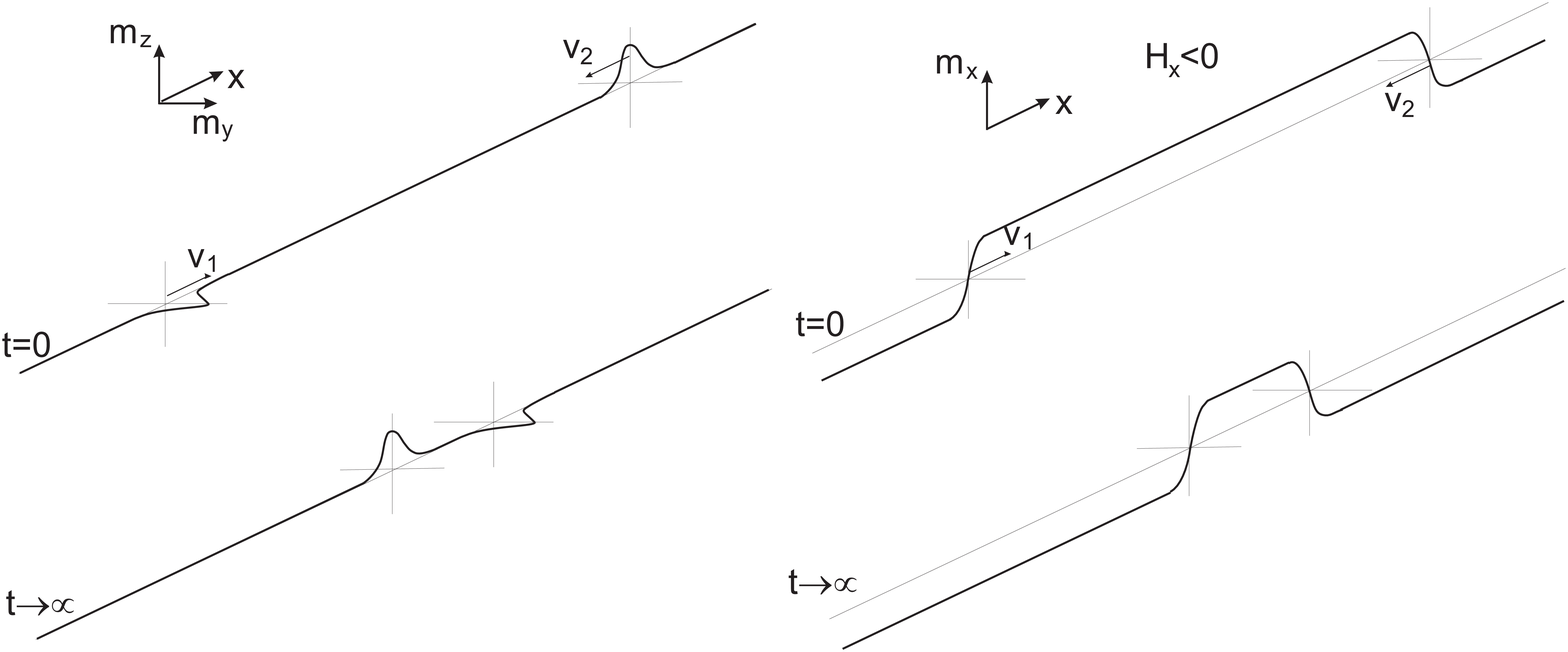}}}
\end{picture}
\end{center}
\caption{Formation of hard bubble (composed of one Neel DW and one Bloch DW) in a longitudinal field. The DW reflection is accompanied by a change of the Bloch wall (of the velocity $v_{2}$) into the Neel wall (of the velocity $v_{1}$) and vice versa.}
\end{figure}
 
In order to complete the picture of the bubble formation in 1D ferromagnet, I refer to my previous study of 
 complexes of one Bloch DW and of one Neel DW \cite{jan11}. In the absence of external field, such a stationary complex 
 is described by a strict solution to the LLG equation, hence, there is no interaction between the walls. 
 An interaction appears, however, upon the field application due to a dynamically induced deformation of the DWs.
 The external field enforces motion of any DW and its anti-wall in the opposite directions and it can induce the collision of a Neel DW 
 with a Bloch DWs. The result of such a collision has been found to be their mutual reflection accompanied by the transformation
 of the Bloch wall into the Neel DW and vice versa. The process can be considered as the transmission
 of the Bloch DW through the Neel DW with change of the head-to-head structure into the tail-to-tail one (and vice versa) which 
 is illustrated in Fig. 2, and it is similar to the collision of spontaneously propagating (in absence of dissipation) DWs \cite{kos90}.
 Since, under the action of the external field, the different-type DWs interact repulsively, they can form 
 a bubble which is an 1D counterpart of the hard bubble in planar magnets. 
 

\section{Bubbles near criticality}

The subcritical dynamics of bistable systems is governed by the GL equation
\begin{eqnarray}
\alpha\frac{\partial m}{\partial t}=J\frac{\partial^{2}m}{\partial x^{2}}
+\beta_{1}m+\beta_{2}m^{*}-\mu|m|^{2}m+\gamma H.
\label{GL}
\end{eqnarray} 
For $\beta_{1}>3\beta_{2}$, (\ref{GL}) describes the evolution of Bloch (of lower energy) and Ising (of higher energy) DWs
 and their complexes. The field-induced creation of 1D hard bubble (composed of one Bloch DW and one Ising DW)
 has been investigated in \cite{jan11a}. It has been found to be similar to the dynamics of the hard bubble
 in the low temperature regime. For $H=0$, the stationary complex of one Ising DW and one Bloch DW 
 in an 1D subcritical system is described by a strict solution to (\ref{GL}), thus, 
 both the DWs do not interact \cite{bar05}. The field-driven collision of the Ising DW with 
 the Bloch DW induces their mutual reflection, hence, they can form a similar hard bubble as it was described in the previous
 section. I mention that an analysis of linear-wave scattering on the Bloch-Ising complex with the parametrically-driven 
 nonlinear Schrodinger equation has shown the dynamically-induced repulsion of both the walls also \cite{bar07a}.
 Below I pay my attention to soft magnetization bubbles (of two Bloch DWs or of two Ising DWs).
 
In the first perturbation approximation, in the vicinity of the center of $j$th DW (j=1,2), the magnetization
 profile can be written as 
\begin{eqnarray} 
m(x,0)=m^{(j)}(x)+\delta m^{(j)}(x),
\end{eqnarray} 
where $m^{(j)}$ denotes a strict (stationary) single-DW solution to (\ref{GL}) for $H=0$
\begin{eqnarray}
m^{(j)}(x)&=&\sqrt{\frac{\beta_{1}+\beta_{2}}{\mu}}{\rm tanh}[\sigma_{j}k(x-x_{0j})]
\nonumber\\
&&+{\rm i}\sin{\varphi_{j}}\sqrt{\frac{\beta_{1}-3\beta_{2}}{\mu}}{\rm sech}[\sigma_{j}k(x-x_{0j})].
\label{j_thDW}
\end{eqnarray}
Here $\sigma_{1}=-\sigma_{2}=\pm 1$, and $k=\sqrt{2\beta_{2}/J}$, $\varphi_{j}=\pm\pi/2$ (Bloch DWs)
 or $k=\sqrt{(\beta_{1}+\beta_{2})/(2J)}$, $\varphi_{j}=0,\pi$ (Ising DWs). 
 According to my claim on the form of the ansatz (\ref{lowest_order_LLG}), since the interaction of DWs is related to 
 a reorientation of the separating them domain, I expect the perturbed magnetization not to contain linear in $m^{(j)}$ 
 terms. Taking the following perturbation $\delta m^{(j)}=m^{(j)}[\pm m^{(k)}\sqrt{\mu/(\beta_{1}+\beta_{2})}-1]$,
 ($k\neq j$), leads to
\begin{eqnarray}
m=\pm\sqrt{\frac{\mu}{\beta_{1}+\beta_{2}}}m^{(1)}m^{(2)}.
\label{lowest_order_GL}
\end{eqnarray} 
This form of the initial magnetization was used in \cite{bar07}, whereas, previous treatments of the DW interactions within
 the GL approximation used the ansatz $\delta m^{(j)}=m^{(k)}\pm\sqrt{(\beta_{1}+\beta_{2})/\mu}$, ($k\neq j$), by an analogy 
 to the perturbation calculus for nontopological (bright) solitons \cite{kor92}.
 
\begin{figure}
\unitlength 1mm
\begin{center}
\begin{picture}(175,30)
\put(0,-5){\resizebox{85mm}{!}{\includegraphics{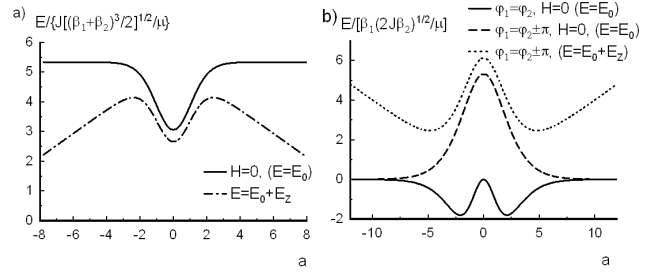}}}
\end{picture}
\end{center}
\caption{Energy of a pair of DWs as a function of the distance of their separation; a) solid (dash-dotted) line - 
Ising DWs in absence (presence) of an external field, b) solid line - Bloch DWs of the opposite chiralities 
(an untwisted pair), dashed (dotted) line - Bloch DWs of the same chiralities (a twisted pair) 
in absence (presence) of an external field.}
\end{figure}
 
Inserting (\ref{lowest_order_GL}) into the Hamiltonian 
\begin{eqnarray}
{\cal H}={\cal H}_{0}+{\cal H}_{Z}=J\bigg|\frac{\partial m}{\partial x}\bigg|^{2}
+\frac{\mu}{2}\left(|m|^{2}-\frac{\beta_{1}+\beta_{2}}{\mu}\right)^{2}
\nonumber\\
-\frac{\beta_{2}}{2}(m-m^{*})^{2}-\gamma H\left(m+m^{*}\pm 2\frac{\sqrt{\beta_{1}+\beta_{2}}}{\sqrt{\mu}}\right),
\end{eqnarray}
one arrives at the dependence of the energy $E_{0(Z)}\equiv\int_{-\infty}^{\infty}{\cal H}_{0(Z)}{\rm d}x$
 on the distance between the DW centers. Defining $a\equiv k(x_{02}-x_{01})$, for the interacting Ising DWs, one finds
\begin{eqnarray}
E_{0}(a)&=&\frac{(\beta_{1}+\beta_{2})^{2}}{2\mu k}I_{4}(a)
\nonumber\\
&=&\frac{\sqrt{J}(\beta_{1}+\beta_{2})^{3/2}}{3\sqrt{2}\mu}
{\rm coth}^{2}(a){\rm csch}^{5}(a)[120a{\rm cosh}(a)
\nonumber\\
&&-80{\rm sinh}(a)
-15{\rm sinh}(3a)+{\rm sinh}(5a)],\\
E_{Z}(a)&=&-\gamma H2\frac{\sqrt{\beta_{1}+\beta_{2}}}{\sqrt{\mu}k}I_{5}(a)
\nonumber\\
&=&-\gamma H4\frac{\sqrt{2J}}{\sqrt{\mu}}a{\rm coth}(a)
\end{eqnarray}
while for the Bloch DWs, ($E_{0}$ is derived up to the lowest order in $\beta_{2}/\beta_{1}$);
\begin{eqnarray}
E_{0}(a)&=&\frac{2\beta_{1}\beta_{2}}{\mu k}\left[I_{1}^{\pm}(a)-I_{2}^{\pm}(a)\right]
\nonumber\\
&=&\frac{\beta_{1}\sqrt{2J\beta_{2}}}{\mu}
2{\rm csch}^{2\mp 1}(a/2)
\nonumber\\
&&\times{\rm sech}^{2\pm 1}(a/2)[-a{\rm cosh}(a)+{\rm sinh}(a)].
\\
E_{Z}(a)&=&-\gamma H2\frac{\sqrt{\beta_{1}+\beta_{2}}}{\sqrt{\mu}k}I_{3}^{\pm}(a)
\nonumber\\
&=&-\gamma H4\frac{\sqrt{J(\beta_{1}+\beta_{2})}}{\sqrt{2\beta_{2}\mu}}a[{\rm coth}(a)\mp{\rm csch}(a)].
\end{eqnarray}
The integrals $I_{k}(a)$ are given in Appendix. Here, the upper signs correspond to a pair of the Bloch walls
 of opposite chiralities, the case $\varphi_{1}=\varphi_{2}$ while the lower signs to a pair
 of the Bloch walls of like chiralities, the case $\varphi_{1}=\varphi_{2}\pm\pi$.
 Plotting the energy $E_{0}(a)$ of the DW pairs (Fig. 3), we see the Ising DWs to attract each other in the absence 
 of external field while the character of the interaction of Bloch DWs to be dependent on their chiralities. 
 In the presence of an external field parallel to the bubble magnetization, the Ising walls can form a stationary
 state corresponding to the maximum of energy, however, it is ustable to perturbations.
 The Bloch DWs of opposite chiralities attract each other when the distance 
 of their separation is big and they repel each other at short separation distances. Because of the
 minims of the energy $E_{0}(a)$ at $a\neq 0$, stable bubble of Bloch DWs (a breather) can be formed even
 in absence of the field. The Bloch DWs of like chiralities always repel in absence of the field, 
 hence, an external field antiparallel to the magnetization of the intermediate domain enables creation
 of a stable bubble (a breather) which corresponds to the minims of $E_{0}(a)+E_{Z}(a)$.
    
\section{Structures of many DWs}

Stability of many-DW structures is an important problem in view of the challenge of designing a nanowire-based information carrier
 (a DW-racetrack memory) \cite{par08}. Periodic structures of the Bloch (Neel) DWs are stationary because of the compensation
 of (attractive or repulsive) interactions between the walls. However, bit recording requires switching the magnetization of a single
 domain of the memory tape on demand. For instance, the magnetization reversal can be performed via flip of the chirality
 of two neighboring DWs in the tape, which initiates their movement towards or outwards each other and, eventually, results in their 
 annihilation due to the appearance of uncompensated attractive forces. Unfortunately, the resulting nonperiodicity of the system 
 causes destabilization of the record. I mention that similar process is responsible for annihilation of DWs during strong-current
 driven motion of multi-domain systems \cite{mur10}. Then, the unbalanced interactions appear because, for the current intensity
 exceeding the Walker-breakdown value, magnetization in different (head-to-head and tail-to-tail) DWs rotates in opposite 
 directions about the magnetic axis, (unlike upon the application of a strong magnetic field). 

An exception is a subcritical system of Bloch DWs whose all the neighboring DWs are of opposite chiralities
 and the distance of their separation corresponds to the size of the breather found in section III. The result of 
 changing the chirality of a single DW or a DW pair is some shift of the wall positions without any flip of the 
 domains.
 
Aperiodic trains of DWs of like chiralities (unstable to DW interactions) as well as trains of alternating Bloch and Neel (Ising)
 DWs (unstable to the Neel-Bloch or Ising-Bloch transition inside the DWs, \cite{laj79}) can be stabilized by an external magnetic field
 whose application results in creation of soft and hard magnetization bubbles, respectively. 
 The energy of a soft bubble increases with the field value [see minims of $E(a)$, dotted lines in Fig. 1 and Fig. 3], hence,
 unlike for weak fields, in a regime of strong field, the hard bubbles can be energetically favorable \cite{ros72}.
      
\section*{Acknowledgements}

This work was supported by Polish Government Research Founds for 2010-2012 in the framework of Grant No. N N202 198039.

\appendix
\begin{widetext}
\section{Explicit form of integrals}

\begin{eqnarray}
I_{1}^{\pm}(a)&\equiv&\int_{-\infty}^{\infty}\Bigl(\left\{{\rm sech}^{2}(y){\rm tanh}(-y+a)
-{\rm sech}^{2}(-y+a){\rm tanh}(y)\mp\left[-{\rm tanh}(y)+{\rm tanh}(-y+a)\right]{\rm sech}(y)
{\rm sech}(-y+a)\right\}^{2}
\Bigr.\nonumber\\
&&\left.
+\left\{-{\rm sech}(y)\left[{\rm tanh}(y){\rm tanh}(-y+a)+{\rm sech}^{2}(-y+a)\right]\pm
{\rm sech}(-y+a)\left[{\rm tanh}(y){\rm tanh}(-y+a)+{\rm sech}^{2}(y)\right]\right\}^{2}\right){\rm d}y
\nonumber\\
&&\\
I_{2}^{\pm}(a)&\equiv&\int_{-\infty}^{\infty}\left(
\left[{\rm sech}(y){\rm tanh}(-y+a)\pm{\rm tanh}(y){\rm sech}(-y+a)\right]^{2}\right){\rm d}y
\\
I_{3}^{\pm}(a)&\equiv&
\int_{-\infty}^{\infty}\left[{\rm tanh}(y){\rm tanh}(-y+a)\mp{\rm sech}(y){\rm sech}(-y+a)+1\right]{\rm d}y
\\
I_{4}(a)&\equiv&
\int_{-\infty}^{\infty}\left\{\left[{\rm sech}^{2}(y){\rm tanh}(-y+a)
-{\rm sech}^{2}(-y+a){\rm tanh}(y)\right]^{2}+\left[{\rm tanh}^{2}(y){\rm tanh}^{2}(-y+a)-1\right]^{2}\right\}{\rm d}y
\\
I_{5}(a)&\equiv&
\int_{-\infty}^{\infty}\left[{\rm tanh}(y){\rm tanh}(-y+a)
+1\right]{\rm d}y
\end{eqnarray}

\end{widetext}

\end{document}